\documentclass{article}

    \usepackage[final]{neurips_2023}

\usepackage[utf8]{inputenc} 
\usepackage[T1]{fontenc}    
\usepackage{hyperref}       
\usepackage{url}            
\usepackage{booktabs}       
\usepackage{amsfonts}       
\usepackage{nicefrac}       
\usepackage{microtype}      
\usepackage{xcolor}         

\usepackage{amsmath}
\usepackage{amssymb}
\usepackage{mathtools}
\usepackage{amsthm}
\usepackage{algorithm2e}
\RestyleAlgo{ruled}
\SetKwInput{KwInput}{Input}                
\SetKwInput{KwOutput}{Output}              
\SetKwComment{Comment}{/* }{ */}
\usepackage[inline]{enumitem}

\usepackage{floatrow}

\usepackage{blindtext}

\usepackage[capitalize,noabbrev]{cleveref}

\title{AttentionStitch: How Attention Solves the Speech Editing Problem}

\author{%
 Antonios Alexos\thanks{corresponding author} \\
  Department of Computer Science\\
  University of California Irvine\\
  \texttt{aalexos@uci.edu} \\
  \And
  Pierre Baldi \\
  Department of Computer Science\\
  University of California Irvine\\
  \texttt{pfbaldi@uci.edu} \\
}

\begin{document}
\maketitle

\begin{abstract}
The generation of natural and high-quality speech from text is a challenging problem in the field of natural language processing. In addition to speech generation, speech editing is also a crucial task, which requires the seamless and unnoticeable integration of edited speech into synthesized speech. We propose a novel approach to speech editing by leveraging a pre-trained text-to-speech (TTS) model, such as FastSpeech 2, and incorporating a double attention block network on top of it to automatically merge the synthesized mel-spectrogram with the mel-spectrogram of the edited text. We refer to this model as AttentionStitch, as it harnesses attention to stitch audio samples together. We evaluate the proposed AttentionStitch model against state-of-the-art baselines on both single and multi-speaker datasets, namely LJSpeech and VCTK. We demonstrate its superior performance through an objective and a subjective evaluation test involving 15 human participants. AttentionStitch is capable of producing high-quality speech, even for words not seen during training, while operating automatically without the need for human intervention. Moreover, AttentionStitch is fast during both training and inference and is able to generate human-sounding edited speech\footnote{Audio samples can be found \href{https://attentionstitch.github.io}{webpage} or \href{https://drive.google.com/drive/folders/1Afc2BzgRqmL_MyyKqXS5RBSR5DGppMWf?usp=share_link}{gdrive link}.}.
\end{abstract}

\section{Introduction}
\label{sec:intro}

TTS synthesis has paved the way for the exploration of various speech tasks like speech editing. In speech editing, the edited part is synthesized and then combined with the rest of the reference audio to produce a smooth and natural-sounding audio sample. The speech editing task can be formally defined as follows: given a reference audio sample $S_R$ with its corresponding text transcript $T_R$, the objective is to modify $T_R$ by replacing word(s), resulting in an edited text transcript $T_E$. $T_E$ is then synthesized into an edited audio sample $S_E$, with the aim of making $S_E$ sound similar to $S_R$ and for the edited portion of $S_E$ to be indistinguishable from the original $S_R$ to the listener. One method for speech editing integrates segments from the same speaker using pitch and prosody features for natural editing \cite{morrison2021context}. Another approach generates audio in a generic voice and converts it to the desired target voice \cite{jin2017voco}, but has noticeable roughness at edit boundaries. EditSpeech \cite{tan2021editspeech} uses forward and backward decoders for fused mel-spectrograms, while A$^3$T \cite{bai20223} introduces cross-modal alignment embedding. EdiTTS \cite{tae2021editts} refines edited speech with perturbations to Gaussian priors. SpeechPainter \cite{borsos2022speechpainter} fills speech gaps using an attention-based model, limited to 1-second gaps. MaskedSpeech \cite{zhang2022maskedspeech} focuses on Mandarin speech editing with a pretrained FastSpeech2 model. Our novel approach expedites training and achieves smoother audio segment integration by incorporating an auxiliary module into a pretrained TTS model like FastSpeech2 (FS2), enhancing efficiency and naturalness in audio stitching. This paper presents a novel method for speech editing that is rooted in state-of-the-art TTS synthesis with the following contributions: \begin{enumerate*}[label=(\arabic*)]
    \item Proposal of AttentionStitch, a speech editing model combining FS2 and a double attention block.
    \item Fast and high-quality synthesis with automatic editing due to attention.
    \item Subjective and objective evaluation on single (LJSpeech) and multi-speaker (VCTK) data, showing superiority over state-of-the-art methods and an extensive ablation study.
\end{enumerate*}

\section{Preliminaries}
\label{sec:prelim}


{\bf Double Attention Block.} The role of the double attention block is to propagate global features from images and enable the model to access them efficiently. It operates in two steps. In the first step, the double attention block gathers image features through an attention-pooling operation. In the second step, it selects and distributes the features via attention. Let $X \in \mathbb R^{c\times d\times h\times w}$ be the input to a 3D convolution layer, where $c$ denotes the number of channels, $d$ the dimension, and $h$ and $w$ the dimensions of the images. In a general format, we define $F_{distr}$ as the operation that distributes the features in the layers, and $F_{gather}$ is the operation that gathers the features from the images which are later distributed. So for every location $i, \dots, dhw$ of the network with feature $u_i$ we have the output of an operation that gathers features from each location $i$ and distributes them back to each location considering the local feature.


{\bf FastSpeech 2.} The core model utilized in our approach is a pre-trained FS2 model, which is a fast-training non-autoregressive TTS synthesis model. FS2 follows a two-step synthesis process, where it initially predicts a mel-spectrogram and then transforms it into an audio waveform. To enhance the naturalness and controllability of the synthesized speech, FS2 incorporates prosody features such as energy, pitch, and duration. These features are predicted individually through dedicated predictors during training, while during inference, they are predicted by the model itself. The model follows an Encoder-Decoder architecture, taking phonemes as input and generating a mel-spectrogram as output. The mel-spectrogram is subsequently processed by a vocoder, which generates the corresponding audio waveform. The Encoder module receives the phoneme embedding and positional embedding as input and produces a hidden sequence. A module called the variance adaptor enriches this hidden sequence with informative features such as energy, pitch, and duration. Finally, the Decoder module parallelly converts the enriched hidden sequence into a mel-spectrogram.


\section{Proposed Method}
\label{sec:prop_meth}

AttentionStitch comprises a pre-trained FS2 model and a double attention block. We chose FS2 since it is renowned as one of the top-performing models in the TTS community, offering fast training and inference speeds with high-quality speech synthesis. The FS2 model is already pre-trained as part of AttentionStitch, enabling us to save time during training. We only need to train the remaining components of the model, which is more manageable due to its smaller size. The double attention block employed is intuitive in its operation and serves the purpose of gathering features from the synthesized mel-spectrogram to fill the masked regions of the reference mel-spectrogram. The application of this model variation in the context of audio data makes it a suitable candidate for achieving the desired combination of features from the synthesized and reference mel-spectrograms. This choice is motivated by the need to combine specific parts of two mel-spectrograms effectively and automatically, which aligns with the objective of synthesizing high-quality speech output. 

\begin{figure}[htbp]
     \RawFloats
    \begin{minipage}{0.55\textwidth}
        \centering
        \includegraphics[width=.6\linewidth]{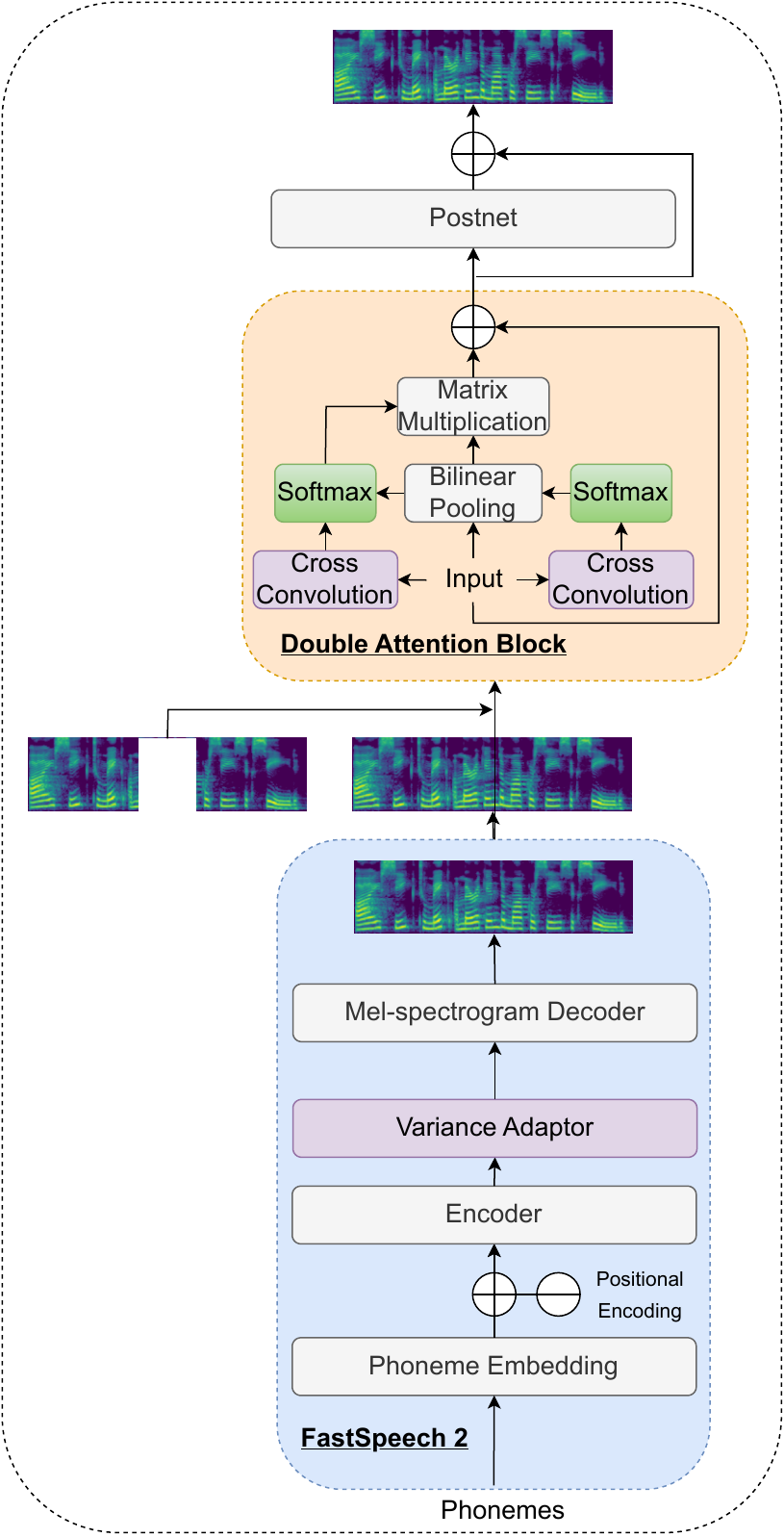}
        \caption{Overview of our proposed AttentionStitch model. AttentionStitch consists of a pre-trained FS2 model and a Double Attention Block.}
        \label{fig:attentionstitch}
    \end{minipage}%
    \begin{minipage}{0.4\textwidth}
        \renewcommand{\arraystretch}{1.2}
        \setlength{\abovecaptionskip}{5pt}
        \begin{tabular}{|c|c|}
            \hline 
            \textbf{Method} & \textbf{MOS ($\uparrow$)} \\\hline
            Complete synthesis & 2.56$\pm$0.33          \\\hline
            FeatSwitch         & 2.73$\pm$0.35               \\\hline
            \textbf{AttentionStitch}    & \textbf{3.86$\pm$0.28}             \\\hline
            Reference          & 4.82$\pm$0.12            \\\hline
        \end{tabular}
        \caption{MOS ($\uparrow$) scores for AttentionStitch, the compared methods, and the reference samples with 95\% confidence intervals for LJSpeech. AttentionStitch outperforms the compared methods.}
        \label{tab:ljspeech_mos_scores}
        
        \hspace{20pt} 

        \setlength{\abovecaptionskip}{5pt}
        \begin{tabular}{|c|c|c|}
            \hline 
            \textbf{Method} & \textbf{MOS ($\uparrow$)} & \textbf{MCD ($\downarrow$)}\\\hline
            EditSpeech & 3.28$\pm$0.33 & 7.54  \\\hline
            A$^3$T & 3.3$\pm$0.35 & 7.97    \\\hline
            \textbf{AttentionStitch} & \textbf{3.51$\pm$0.23} & \textbf{6.5}\\\hline
            Reference  & 4.43$\pm$0.2 & - \\\hline
        \end{tabular}
        \caption{MOS ($\uparrow$) and MCD ($\downarrow$) scores for AttentionStitch, the compared methods, and the reference samples with 95\% confidence intervals for VCTK. AttentionStitch outperforms the compared methods in both metrics.}
        \label{tab:vctk_scores}
    \end{minipage}
\end{figure}

FS2 takes the phonemes of the edited text as input and generates a mel-spectrogram as output. During the training phase, we randomly mask 10\% of the reference mel-spectrogram near its center, although the procedure works if we mask in the beginning or in the end instead. The mask consists of zeros, and we provide further details on the masking strategy in \cref{subsec:multi_speaker_results}. After masking the reference mel-spectrogram, we concatenate it with the synthesized mel-spectrogram and feed it into the double attention block which redistributes the features of the synthesized mel-spectrogram within the masked region of the reference mel-spectrogram. Additionally, we employ a Postnet module to refine the mel-spectrogram further. Finally, a Hifi-GAN \cite{kong2020hifi} vocoder, transforms the final mel-spectrogram into an audio waveform. We incorporate skip connections between the output of the double attention block and the Postnet, as they have proven to be beneficial components in speech synthesis \cite{tu2017speech, shi2018speech}. During inference we mask the phoneme sequence based on the word(s) that need to be edited, as we know their boundaries. As in training, we perform the masking operation by replacing the corresponding part of the mel-spectrogram with zeros. Additionally, we modify the reference text $T_R$ by replacing the word(s) with the target word(s). To ensure the reference mel-spectrogram and the synthesized mel-spectrogram have the same length, we leverage the duration predictor of FS2 and resize the mask accordingly. The speech editing operation takes place within the double attention block. The proposed AttentionStitch model is depicted in \cref{fig:attentionstitch}.


\section{Experiments and Evaluation}
\label{sec:exp_and_eval}


\subsection{Experimental Setup}
\label{subsec:experiments}

We utilized the experimental setup described in the corresponding papers for the pre-trained FS2 model and double attention network \cite{fs2_torch_impl,double_attention_impl}. Our results were obtained for single and multi-speaker data on the LJSpeech dataset \cite{ljspeech17} and the VCTK dataset \cite{Yamagishi2019} respectively. We pre-trained FS2 for 200,000 steps on both LJSpeech and VCTK. In an effort to improve the performance of the model for speech editing task, we also attempted training the FS2 for 900,000 steps, but observed that the model overfit and performed poorly for this task. We then froze the entire FS2 model and in the second training phase, we trained the double attention block along with a postnet for 200,000 steps. The double attention block was utilized for speech editing, while the postnet was used for better refinement of the mel-spectrogram. The postnet consisted of 5 1-D convolutions with 512 channels and a kernel size of 5. We also observed that high-quality audio samples could be produced with limited training steps. We use Mean Average Error (MAE) as the loss for the double attention block and the postnet loss, while the rest of the model is frozen.

We first evaluate AttentionStitch using a single speaker dataset, LJSpeech, against two baselines derived from the pre-trained FS2 model denoted as ``FeatSwitch" and ``Complete synthesis and swap" which do not require any additional tuning or modules for speech editing. Subsequently, we extend our evaluation to VCTK, a more challenging multi-speaker dataset, where we compare to state-of-the-art methods.

{\bf Complete synthesis and swap.} We first use a pretrained FS2 model to synthesize speech corresponding to the edited text $T_e$. We predict the durations from a pretrained FS2 and use them to find the word boundaries. Then we replace the source word(s) in the reference with the synthesized word(s).

{\bf FeatSwitch.} Within FeatSwitch, we perform mid-inference prosody feature switching in FS2 for speech editing. We extract phoneme-level energy, pitch, and duration features from the reference audio $R$ and predict the same features from edited text $T_e$ using FS2. We then replace ground truth features in all non-target phonemes in the synthesized audio. 

We deliberately chose not to compare with fully resynthesized edited text, since the edited part may be indistinguishable, but the final audio sample still differs from the reference audio. For VCTK, we compared AttentionStitch to more competitive and state-of-the-art methods, namely A$^3$T and EditSpeech. Although other modern methods exist like SpeechPainter, EdiTTS, and MaskedSpeech, we chose not to compare with these based on the nature of these works. SpeechPainter fills limited gaps of audio samples with the same text; EdiTTS is evaluated only on single-speaker data; and MaskedSpeech works on sentence level (not word level) and is evaluated only in Mandarin.

\subsection{Evaluation Setup}
\label{subsec:evaluation}

We evaluated AttentionStitch using the objective metric Mel-cepstral distance (MCD) \cite{kubichek1993mel}, and the subjective metric Mean Opinion Scores (MOS) \cite{sector2017recommendation}. MCD measures the difference between two mel-spectrograms, and MOS on the other hand, is a score given by human listeners based on audio quality. It is worth noting that a higher MOS score and a lower MCD score indicate better quality of speech. 15 subjects participated in our subjective evaluation study and they could listen to the audio samples as many times as they want. For both datasets we used 2 tests to obtain the MOS scores, where the first one is that the subjects have to give their quality opinion score between 1 and 5, about audio samples of different methods, and in the second they have to give their opinion about audio samples of AttentionStitch which we edited with unseen words. For the first part we chose 3 samples for each method for LJSpeech and 4 samples for each method for VCTK. For the second part we chose 12 samples of AttentionStitch for each dataset.

\subsection{Results on single speaker data}
\label{subsec:single_speaker_results}

In the single-speaker part of the evaluation, the subjects scored the samples based on the audio quality and general preference. The results of this part can be found in Table \ref{tab:ljspeech_mos_scores}, which presents the MOS scores with 95\% confidence intervals for each method and the reference audio samples. We observe that AttentionStitch achieved a high MOS score, indicating a high quality of edited synthesized speech. It significantly outperformed the two baselines, although it had a wide range of reported results. Specifically, FeatSwitch achieved a MOS score of 2.73, Complete Synthesis achieved 2.56, AttentionStitch achieved 3.87, and the reference audio samples achieved 4.82. Although the gap between AttentionStitch and the reference seems large, the subjective evaluation results in speech synthesis can vary widely and are highly dependent on individual preferences.

\subsection{Results on multi-speaker data}
\label{subsec:multi_speaker_results}

In the multi-speaker evaluation, following the same methodology as in \cref{subsec:single_speaker_results} and in \cref{subsec:evaluation}, we collected MOS scores from the participants to assess their overall preference and audio quality for each method and reference audio sample. The MOS scores with a 95\% confidence interval for each method and reference audio sample are presented in \cref{tab:vctk_scores}. AttentionStitch achieved the highest MOS score, indicating a superior quality of the synthesized speech. Specifically, AttentionStitch obtained a MOS score of 3.51, while EditSpeech and A$^3$T achieved scores of 3.28 and 3.3, respectively. In contrast, the reference audio samples received a MOS score of 4.43. It is worth noting that the VCTK dataset is more challenging than the LJSpeech dataset due to the variety of accents present, resulting in lower MOS scores for both AttentionStitch and the reference audio samples. We also present our findings with a short objective evaluation with MCD scores which show that our method performs better than A$^3$T and EditSpeech.

As part of an ablation study, we subjected AttentionStitch to the task of synthesizing edited audio containing unseen words; words not encountered during the model's training phase. Remarkably, the resulting MOS obtained from this exercise stood at \textbf{3.66$\pm$0.17}, a value closely aligned with the original MOS reported in \cref{tab:vctk_scores}. Furthermore, we explored the feasibility of altering multiple words within a sentence simultaneously. This endeavor presented challenges, as the model's training involved applying the mask to only one part at a time. Sequential word changes led to a decline in audio quality due to the emergence of electronic artifacts. Nonetheless, AttentionStitch retains the capability to replace a single word with multiple words, showcasing its versatility in handling certain editing tasks.

\section{Conclusions and Discussion}
\label{sec:conclusions}

AttentionStitch is a novel method for speech editing that utilizes a pre-trained FS2 model and incorporates the unique double attention block which effectively gathers the features of the edited part and distributes them within the masked area of the reference mel-spectrogram. It is a fast approach to speech editing for researchers with limited resources, something that the community has not addressed yet.

\bibliography{bibliography}
\bibliographystyle{bibliography}


\end{document}